\documentclass[aps,prl,twocolumn,amsmath,amssymb,nofootinbib,showpacs,superscriptaddress]{revtex4-1}
\usepackage[english]{babel}
\usepackage{latexsym}
\usepackage{graphics}
\usepackage{graphicx}
\usepackage{epsfig}
\usepackage{color}
\usepackage{bm}
\usepackage{amsmath}
\usepackage{amssymb}
\usepackage{amsthm}
\usepackage{dcolumn}
\usepackage{bm}
\usepackage{float}
\usepackage{hyperref}
\usepackage{color}
\usepackage{epstopdf}
\usepackage{cleveref}
\usepackage[svgnames]{xcolor}
\hypersetup{hidelinks,colorlinks=true,allcolors=DarkBlue}
\usepackage{enumerate}

\begin{document}

\preprint{APS/123-QED}

\title{Demonstration of a quantum key distribution network \\ in urban fibre-optic communication lines}

\author{E.O.~Kiktenko}
\affiliation{Russian Quantum Center, Skolkovo, Moscow 143025, Russia}
\affiliation{Steklov Mathematical Institute of Russian Academy of Sciences, Moscow 119991, Russia}
\affiliation{Bauman Moscow State Technical University, Moscow 105005, Russia}
\author{N.O.~Pozhar}
\affiliation{Russian Quantum Center, Skolkovo, Moscow 143025, Russia}
\affiliation{Bauman Moscow State Technical University, Moscow 105005, Russia}
\author{A.V.~Duplinskiy}
\affiliation{Russian Quantum Center, Skolkovo, Moscow 143025, Russia}
\affiliation{Moscow Institute of Physics and Technology, Dolgoprudny, Moscow Region 141700, Russia}
\author{A.A.~Kanapin}
\affiliation{Russian Quantum Center, Skolkovo, Moscow 143025, Russia}
\affiliation{Lomonosov Moscow State University, Moscow 119992, Russia}
\author{A.S.~Sokolov} 
\affiliation{Russian Quantum Center, Skolkovo, Moscow 143025, Russia}
\author{S.S.~Vorobey}
\affiliation{Russian Quantum Center, Skolkovo, Moscow 143025, Russia}
\author{A.V.~Miller} 
\affiliation{Russian Quantum Center, Skolkovo, Moscow 143025, Russia}
\author{V.E.~Ustimchik}
\affiliation{Russian Quantum Center, Skolkovo, Moscow 143025, Russia}
\author{M.N.~Anufriev}
\affiliation{Russian Quantum Center, Skolkovo, Moscow 143025, Russia}
\affiliation{Bauman Moscow State Technical University, Moscow 105005, Russia}
\author{A.S.~Trushechkin}
\affiliation{Steklov Mathematical Institute of Russian Academy of Sciences, Moscow 119991, Russia}
\author{R.R.~Yunusov}
\affiliation{Russian Quantum Center, Skolkovo, Moscow 143025, Russia}
\author{V.L.~Kurochkin}
\affiliation{Russian Quantum Center, Skolkovo, Moscow 143025, Russia}
\author{Y.V.~Kurochkin}
\affiliation{Russian Quantum Center, Skolkovo, Moscow 143025, Russia}
\author{A.K.~Fedorov}\email{akf@rqc.ru}
\affiliation{Russian Quantum Center, Skolkovo, Moscow 143025, Russia}

\begin{abstract}
We report the results of the implementation of a quantum key distribution (QKD) network using standard fibre communication lines in Moscow.\,The developed QKD network is based 
on the paradigm of trusted repeaters and allows a common secret key to be generated between users via an intermediate trusted node. 
The main feature of the network is the integration of the setups using two types of encoding, i.e. polarisation encoding and phase encoding. 
One of the possible applications of the developed QKD network is the continuous key renewal in existing symmetric encryption devices with a key refresh time of up to 14 s.
\end{abstract}

\maketitle

\section{Introduction}

During last decades, significant progress in theory, experimental study, and technology of QKD has been performed~\cite{Gisin2002,Lo2015,Lo2016}. 
However, QKD technology faces a number of challenges such as distance, key generation rate, practical security, and others~\cite{Gisin2002,Lo2015,Lo2016}.
In order to make QKD technology available for multiple users, QKD networks are required~\cite{Laenger2009}. 
There is a number of large projects on creating QKD networks, in particular, in United States, Europe, China, 
and Japan~\cite{Yeh2005,Peev2009,Stucki2011,Pan2009,Pan2010,Han2010,Zeilinger2011,Shields2013,Zhang2016}. 
QKD networks have a number of promising applications, for example, the development of secure distributed databases~\cite{Kiktenko20172}.
First of all, they offer information-theoretic secure communications between nodes.
Moreover, the generated keys can be used for continuous key renewal in the currently available symmetric cipher devices.

One of the most important challenges in the development of QKD networks is establishing secret keys beyond laboratory conditions~\cite{Sokolov2017}.
Thereby it is important to use a QKD protocol, that guarantees secrecy in urban fibers with significant losses.
This circumstance is one of the most important distinguishing factors of experiments on the quantum key distribution in urban conditions. 
It is also important to note that the post-processing procedures of sifted keys are an inherent part of QKD networks~\cite{Kiktenko2016}.

The purpose of this work is an experimental demonstration of QKD networks for systems with different types of quantum-state encoding in urban conditions. 
The quantum key distribution is implemented using a `dark', high-loss optical fibre, which is laid together with the available communication lines. 
One of possible applications of the developed QKD network is the continuous key renewal in the currently available symmetric encryption devices. 
Russian encryption standards assume the use of keys with a length of 256 bits, and therefore, taking into account the use of QKD networks, they can be refreshed approximately every 14 s.

\begin{figure}[t]
\begin{centering}
\includegraphics[width=1\columnwidth]{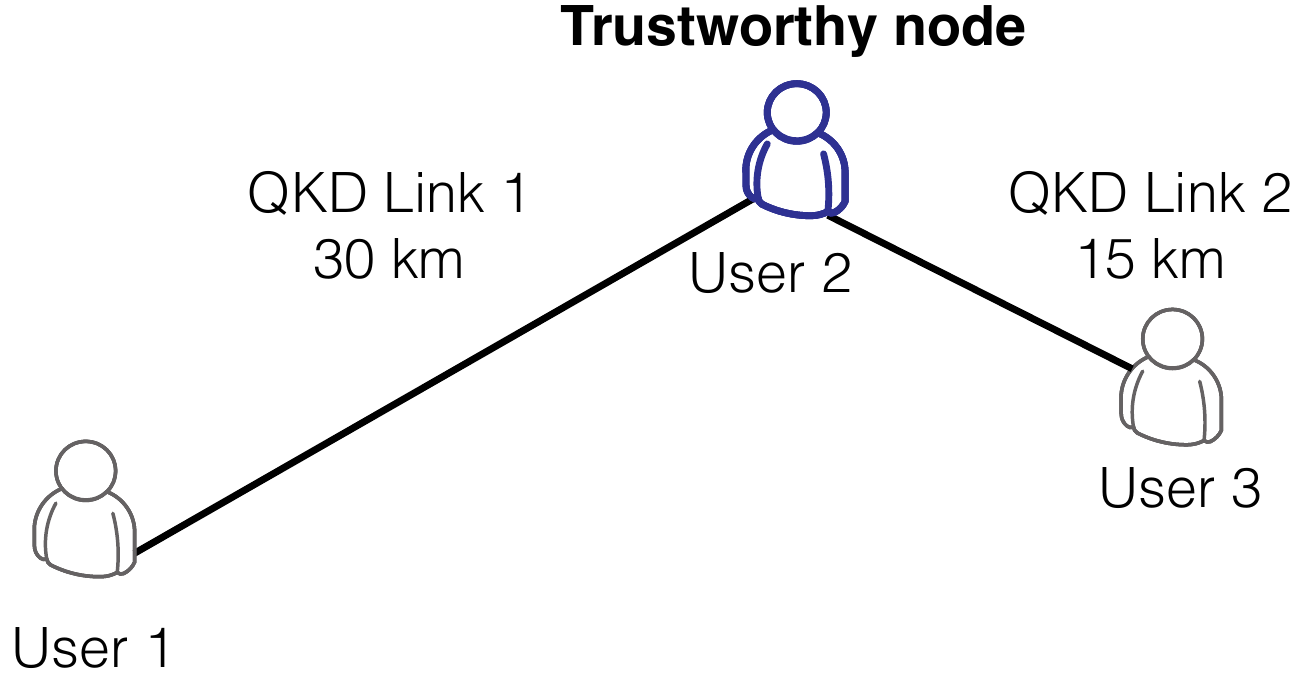}
\end{centering}
\vskip -6mm
\caption
{Quantum keys transport between three users over an intermediate trusted node. The first link generates quantum keys using the polarization-encoding scheme, whereas the second link employs the phase-encoding scheme.}
\label{fig:topology}
\end{figure}

\section{QKD network}

\begin{figure*}
\begin{center}
\includegraphics[width=1\linewidth]{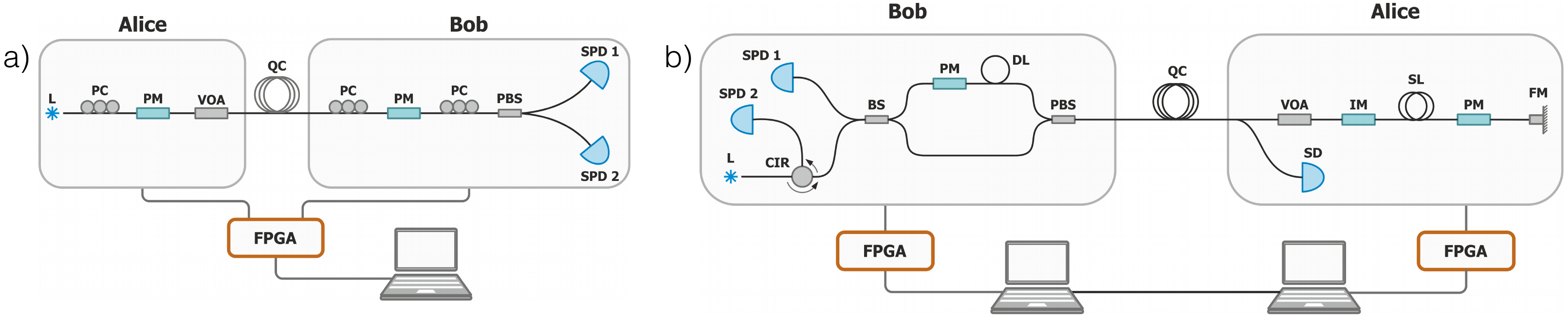}
\vskip -4mm
\caption
{In (a) the first QKD link for generation quantum keys using the polarization-encoding scheme is presented, 
where L is the light source, PC is the polarization controller, PM is the phase modulator, VOA is the variable optical attenuator, QC is the quantum channel (urban fibre channel), PBS is the polzarization beamsplitter, and SPD is the single-photon detector.
In (b) the second QKD link employing the phase-encoding scheme is presented, 
where CIR is the circulator, BS is the beamsplitter, DL is the delay line, SL is the storage line, SD is the synchro detector, IM is the intensity modulator, and FM is the Faraday mirror.}
\label{fig:setup}
\end{center}
\end{figure*}

We employ the SECOQC approach~\cite{Peev2009}, which defines a QKD network as an infrastructure based on point-to-point QKD capabilities.
Then any two nodes of the QKD network can establish a common private key using unconditionally secure transport.
The network protocol in our case --- three nodes and two QKD links --- works as follows. 
Node 1 and Node 2 as well as Node 2 and Node 3 establish their secret keys $k_{12}$ and $k_{23}$.
These keys are stored in the memory of corresponding nodes.
Using a quantum random number generator, Node 1 generates a key $K$ and then forward it encrypted by one-time pad,
$K\oplus{k_{12}}$, to the intermediate trustworthy node (Node 2).
Using previously established key $k_{23}$, Node 2 transmits $K\oplus{k_{23}}$ to Node 3.
Therefore, arbitrary nodes (and all nodes together) in the QKD network can establish a common secret key.
We note that in order to ensure that received keys were sent by the particular node, 
information-theoretic secure authentication can be used~\cite{Peev2009}. 

The developed QKD network allows one to establish a common quantum key for users with different QKD optical layouts: 
polarization-encoding and phase-encoding setups.  
The basis for our experimental work is the recently presented modular QKD device~\cite{Sokolov2017}.
This modular QKD device is driven by NI cards with open source LabView code for control and operating, 
open source Python code for the post processing, and open source protocol for external applications~\cite{Kiktenko2016,Kiktenko2017,KiktenkoTrushechkin2016}. 
The QKD setup can operate with any type of single-photon detectors. 
The external drivers of single-photon detectors, phase modulators, and synchronization detectors are realized as removable modules. 
Each device can drive up to 4 detectors and 6 universal ports for lasers, phase or amplitude modulators. 
The software solution in charge of controlling the system is written with the use of the LabVIEW environment. 

Overall control of the electro-optic components is realized by NI PCIe-7811R installed in personal computers~\cite{Sokolov2017}. 
The semiconductor laser LDI-DFB2.5G under the control of FPGA board Spartan-6 
generates 10 MHz frequency optical pulses at the standard telecommunication wavelength 1.55  $\mu$m. 
We use ID230 single-photon detectors~\cite{IDQ}. 
Beamsplitters, Faraday mirror, circulators, variable optical attenuator, phase and intensity modulators are standard optical components. 

The first link of the developed QKD network generates quantum keys using the polarization-encoding scheme on the basis of the BB84 protocol~\cite{BennetBrassard1984}.
In this setup, we have used the Pockels effect of low voltage electro-optical phase modulators based on LiNbO3 (see Fig.~\ref{fig:setup}a). 
We note that this method allows one to employ single laser source for polarization-encoding implementation, 
while most polarization state implementations suffer from the pulse indistinguishability problem~\cite{Duplinskiy2016}.
Furthermore, only two single-photon detectors are required in contrast to the standard polarization schemes with four detectors.
This link ensures the exchange of keys at a distance of up to 30 km (in urban fibers with losses on the level of 13 dB, the mean number of photons is $\mu_{\rm pol}=0.02$) 
with the sifted key generation rate of about 0.1 Kbit/s.

The second link relies on another optical layout for QKD --- the phase-encoding scheme (see Fig.~\ref{fig:setup}b) with the use of the seminal BB84 protocol. 
This scheme has been already used for QKD across urban fiber channels~\cite{Kurochkin2016}.
This link allows one to generate secret keys at a distance of up to 15 km (in urban fibers with losses on the level of 7 dB, the mean number of photons is $\mu_{\rm ph}=0.03$), 
with the sifted key generation rate of approximately 0.2 Kbit/s.

\begin{figure*}
\begin{center}
\includegraphics[width=1\linewidth]{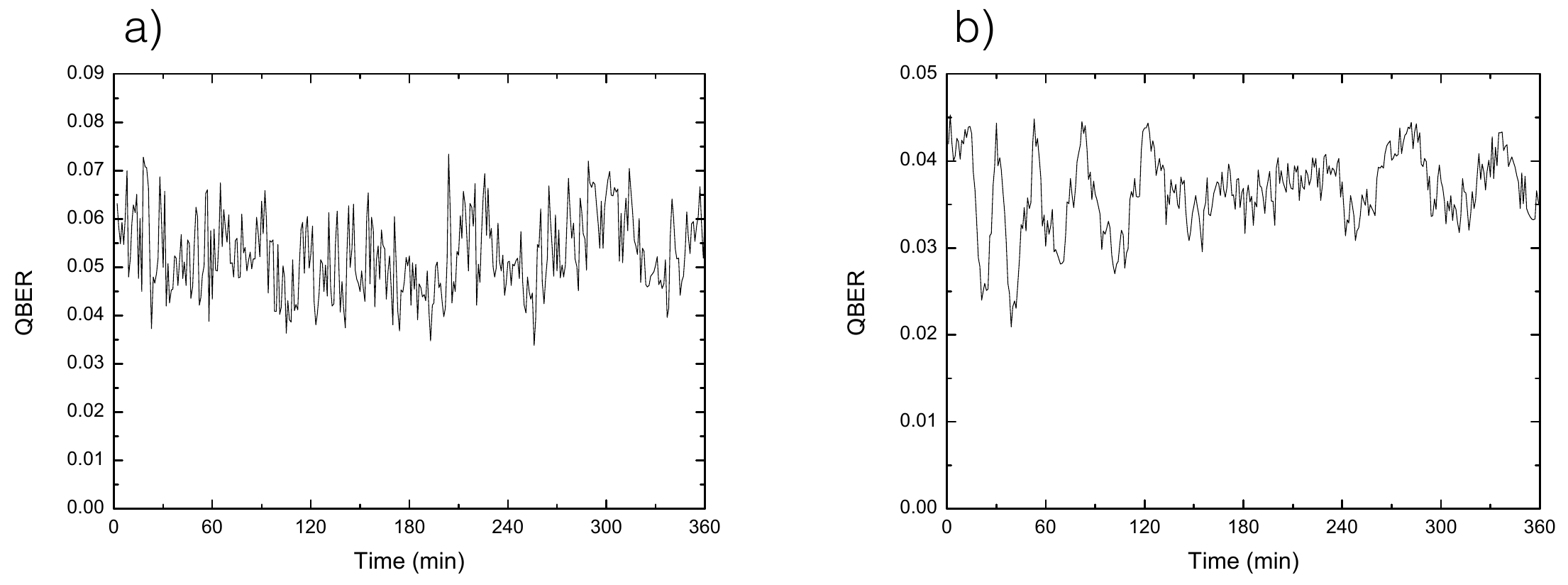}
\vskip -4mm
\caption
{The average value of QBER is shown as the function of time (during 6 hours): (a) for the polarization-encoding setup and (b) for the phase-encoding setup.}
\label{fig:QBER}
\end{center}
\end{figure*}

In Fig.~\ref{fig:QBER} we present the average value of QBER as the function of time (data is presented for 6 hours).
One can see that realistic error rates in the sifted key are of the order of a few percent, which is too high for direct applications, 
e.g., for using as keys for one-time pad encryption or for key renewal in symmetric ciphers.
In order to correct down this error rate as well as reduce this potential information of an eavesdropper to a negligible quantity we use the post-processing procedure, which is described below.

\section{Secret key generation rate}

The sifted keys from both QKD links are the input for the post-processing procedure,
which consists of information reconciliation, parameter estimation, privacy amplification, and authentication check stages~\cite{Kiktenko2016}. 
The procedure works as follows.

\begin{enumerate}[(i)]

\item Sifted keys from the QKD links go through the information reconciliation stage based on the
symmetric blind technique~\cite{Kiktenko2017} based on the use of low-density parity-check codes~\cite{Gallager1962,MacKay1999}.
We note that this technique allows performing an information reconciliation with very rough or even absent estimation on the QBER.  
Furthermore, the blind information reconciliation protocol allows significant increase the efficiency of the error correction procedure and reduces its interactivity~\cite{Kiktenko2017}.

\item After the information reconciliation stage, there is still a certain probability that not all errors are corrected. 
To detect possible remaining errors, we implement a subsequent verification protocol using universal hash functions~\cite{KiktenkoTrushechkin2017}
The probability of the presence of errors after successful verification of the block of $\approx{1}$ Mbit key is limited to the value of $\varepsilon_{\rm ver}=2\times 10^{-11}$ 
when using hash-tag of 50 bit length.
The detailed description of the verification protocol is presented in Ref.~\cite{KiktenkoTrushechkin2017}.
A general binary sequence obtained at this stage is called a \textit{verified key}.

\item In the parameter estimation stage the parties obtain the actual level of the QBER for their key blocks via direct comparison of the keys before and after the information reconciliation.
In fact, this step is performed on the Bob's side where the modification of the sifted key was performed on the previous stage.
If the value of QBER appeared to be higher than the critical value needed for efficient privacy amplification, the legitimate parties abort the protocol. 
Otherwise, the verified key blocks go to the privacy amplification stage, and estimated QBER is used in next rounds of the information reconciliation~\cite{Kiktenko2016}. 

\item The privacy amplification stage is used to reduce potential information of an adversary about the verified blocks to a negligible quantity. 
This is achieved by a contraction of the input verified key into a shorter string. 
The length of the secret key is given by an expression
\begin{equation} \label{eq:pa}
  L_{\rm sec}=L_{\rm ver}\hat{Y}_1(1-h(\hat{q}_1))-{\rm leak}_{\rm ec}-5\log_2(1/\varepsilon_{\rm pa}),
\end{equation}
where $L_{\rm ver}$ is length of the verified key, $\hat{Y}_1$ is an estimation of the portion of the sifted key bits generated from single photons pulses,
\begin{equation}
  h(q)=-q\log_2 q-(1-q)\log_2(1-q)
\end{equation}
is binary entropy function, $\hat{q}_1$ is an estimation of the QBER for single photon pulses, 
${\rm leak}_{\rm ec}$ is total number of bits disclosed in information reconciliation and verification stages, 
and
$\varepsilon_{\rm pa}$ is the failure probability of privacy amplification stages.
In our setup we have adopted the value $\varepsilon_{\rm pa}=10^{-12}$.

The estimation for $\hat{Y}_1$ is as follows:
\begin{equation} \label{eq:Y1}
	\hat{Y}_1 = \frac{\eta\mu-p_2}{\eta\mu},
\end{equation}
where $\eta$ is the transmission coefficient of the quantum channel, $\mu$ is the intensity of the employed laser pulses, 
$p_2=e^{-\mu}\mu^2/2$ is probability of two photons emission in the generation of coherent pulses.
This estimate is obtained on the assumption that Eve can perform an attack with a separation of the number of photons, 
as well as other operations with quantum states being transmitted, but it cannot influence Alice's and Bob's setups 
(for instance, it cannot modify the intensity of the signals sent by Alice or the efficiency of Bob's detector).
In Eq.~\eqref{eq:Y1} we have also neglected the probability of emitting signal with $n>2$ photons.
This assumption is reasonable since our QKD setups use pulses of very low intensity ($\mu_{\rm pol}=0.02$ and $\mu_{\rm ph}=0.03$).
The estimation of the QBER in the single photon pulses, in the assumption that all the errors appeared in single-photon pulses only, is given by the following expression:
\begin{equation} \label{eq:q1}
	\hat{q}_1 = \frac{q}{Y_1},
\end{equation}
where $q$ is the QBER value obtained in parameter estimation step.

After the calculation of the length of a final key (for each verified block of key) according to the method presented in Ref.~\cite{Kiktenko2016}, the privacy amplification can be performed: 
the block of the secret key is computed as a result of application of 2-universal hash function to the verified key.  
In our post-processing procedure the Toeplitz hashing is employed~\cite{Krawczyk1994,Krawczyk1995}.
The resulting key is called the \textit{secret key} or the \textit{final key}. 
It is the output of a QKD protocol.

\item Finally, the parties check the authenticity of their classical channel by exchange of hash values of all the incoming traffic.
In our setup we use Toeplitz hashing together with the one time-pad encryption.
The length of the hash value was set to be $l_{\rm auth}=40$ bit which bounds the probability of successful man-in-the-middle attack at the level of
\begin{equation}
	\varepsilon_{\rm auth}=2\times 2^{-l_{\rm auth}}<2\times 10^{-12}.
\end{equation}

If the authenticity is verified, the parties reserve $2l_{\rm auth}$ bits of their secret keys for the next post-processing round.
Then we arrive at the following expression: 
\begin{equation}
	L_{\rm sec} = L_{\rm sec}-2l_{\rm auth},
\end{equation}
where $L_{\rm sec}$ bits of the final key that can be used in cryptographic purposes.
\end{enumerate}

The final security level of the obtained key is as follows:
\begin{equation}
	\varepsilon_{\rm QKD}=\varepsilon_{\rm ver}+\varepsilon_{\rm pa}+\varepsilon_{\rm auth}<2.3\times 10^{-11}.
\end{equation}
We note that the security level of secret key distributed across the QKD network with  $N$ nodes is given by the following expression:
\begin{equation}
	\varepsilon_{\rm QKDNet}=(N-1)\times (\varepsilon_{\rm QKD}+\varepsilon_{\rm auth}).
\end{equation} 
Here, the term $\varepsilon_{\rm auth}$ comes from the fact that the additional authentication is required.
For our QKD network with $N=3$ we then have $\varepsilon_{\rm QKDNet} < 5\times 10^{-11}$.

If $\tau$ is the time needed to generate a verified key with the length $L_{\rm ver}$, 
then the secret key rate can be defined as follows:
\begin{equation}
	R_{\rm sec}=L_{\rm sec}/\tau.
\end{equation} 
Applying our post-processing procedure to experimentally generated keys, we obtain that the first QKD link provides key exchange over 30 km with the secret key generation rate of about 0.02 Kbit/s.
The second QKD link allows establishing of the secret keys over 15 km, with the secret key generation rate of approximately 0.1 Kbit/s.
The primary application of such quantum keys is continuous key renewal in the currently available symmetric ciphers with the key refresh time up to 14 seconds.

The main application of quantum-distributed keys is the continuous key renewal in the currently available symmetric encryption devices. 
Russian encryption standards assume the use of keys with a length of 256 bits, and therefore, taking into account the use of QKD networks, these keys can be refreshed approximately every 14 s. 
This refresh period is limited by the rate of key generation.

\section{Conclusion}

We have described in detail the implemented QKD networks based on the trusted repeater's paradigm. 
The developed QKD network has been tested using standard fibre-optic communication lines in Moscow. 
It is important to note that the network connects users with two different optical schemes of phase and polarisation encoding.

In designing the network, we have used a 'dark' fibre, laid jointly with the communication lines in use, which produce parasitic illumination at the telecommunication wavelength. 
One of the network links represents a device based on a one-way scheme of key distribution. 
This scheme, in contrast to an auto-compensating one, allows one to send continuous sequences of pulses; 
however, it requires stabilisation relative to the fluctuation of the polarisation state in the quantum channel, caused by external factors (mechanical and thermal effects). 
Unlike the laboratory conditions, tests in a real urban communication line require regular adjustment of parameters and calibration. 
The tests confirmed the capability of the system to compensate for external impacts in conditions of real urban communication lines~\cite{Duplinskiy2016}, 
which allows, in the future, these devices to be integrated into the existing infrastructure.

The main application of such quantum keys is the continuous key renewal in currently available symmetric encryption devices.

The work was supported by the Russian Science Foundation (Grant No. 17-71-20146).

\end{document}